\documentclass[a4paper,floatfix,twocolumn,prb]{revtex4}
\usepackage{graphicx}  

\begin{document}

\title{Broken symmetry and the variation of critical properties in the phase behaviour of supramolecular rhombus tilings}

\author{Andrew~Stannard$^1$, James~C.~Russell$^1$, Matthew~O.~Blunt$^1$, Christos~Salesiotis$^2$, Mar\'{i}a~del~Carmen~Gim\'{e}nez-L\'{o}pez$^2$, Nassiba Taleb$^2$, Martin~Schr\"{o}der$^2$, Neil~R.~Champness$^2$, Juan~P.~Garrahan$^1$ and Peter~H.~Beton$^{1*}$}
\affiliation{$^1$School of Physics and Astronomy, University of Nottingham, University Park, Nottingham, NG7 2RD, UK. 
$^2$School of Chemistry, University of Nottingham, University Park, Nottingham, NG7 2RD, UK. $^*$E-mail: peter.beton@nottingham.ac.uk}

\begin{abstract}
The degree of randomness, or partial order, present in two-dimensional supramolecular arrays of isophthalate tetracarboxylic acids is shown to vary due to subtle chemical changes such as the choice of solvent or small differences in molecular dimensions. This variation may be quantified using an order parameter and reveals a novel phase behaviour including random tiling with varying critical properties as well as ordered phases dominated by either parallel or non-parallel alignment of neighbouring molecules, consistent with long-standing theoretical studies. The balance between order and randomness is driven by small differences in the intermolecular interaction energies, which we show, using numerical simulations, can be related to the measured order parameter. Significant variations occur even when the energy difference is much less than the thermal energy highlighting the delicate balance between entropic and energetic effects in complex self-assembly processes.
\end{abstract}

\maketitle

Two-dimensional molecular networks provide an attractive and highly flexible route to the formation of surfaces with specific structural arrangements and functionalities \cite{Elemans2009,Bartels2010}. Within the broad range of structures that have been explored, random molecular networks have gained considerable attention recently as exemplars of two-dimensional, nonperiodic glassy systems \cite{Blunt2008Science, Otero2008, Marschall2010,Zhou2007}. Although these systems are of great interest, and represent examples of complex self-assembled sytems, a quantitative measure of the degree of randomness present in a molecular array, which is required for a theoretical understanding of their properties, is not generally available. In this paper we report a set of supramolecular arrays in which the randomness varies due to small changes in chemical environment or molecular geometry. We define an order parameter which quantifies these variations and provides a systematic comparison between supramolecular arrays prepared using similar methodologies. Furthermore, we show that the observed variations may be rationalised within a complex phase diagram which can be related to the adsorption of dimers on a periodic lattice\cite{Fisher1961, Kasteleyn1963}, a paradigmatical problem in statistical mechanics which has been studied for several decades, and is a well-known example of a physical system which can be mapped onto a non-periodic tiling of the plane\cite{Henley1999, Destainville2002, Wilson2004, Lu2007, Alet2006, Fradkin2007, Castelnovo2007, Jacobsen2009}. In particular, using numerical simulations, we relate the order parameter to an orientational dependence of the intermolecular interactions which stabilise the network.

We focus on molecular rhombus tilings formed at the liquid-solid interface following the deposition of solvated molecules of interest on highly-oriented pyrolytic graphite (HOPG) substrates. In recent work we have shown\cite{Blunt2008Science} that such arrays support the formation of a random tiling, the first experimental realisation of the dimer adsorption problem. The molecular networks formed on graphite are imaged using scanning tunnelling microscopy (STM) and the observed structures can be directly mapped onto rhombus tilings. In-plane stabilisation of the molecular networks arises from hydrogen bonds formed between carboxylic acid groups \cite{Lackinger2009,Zhou2007,Zhao2010}. For the tetracarboxylic acids considered here this results in two orientations of intermolecular bonding with the molecular backbone axes either parallel or non-parallel at $60^{\circ}$, as shown in Fig.~1a. To form a random tiling a close match between the two dimensions marked $d_1$ and $d_2$ in Fig.~1a is required. Examples of suitable molecules are {\it p}-terphenyl-3,5,3",5"-tetracarboxylic acid (TPTC) (Fig.~1j), with $d_1=8.7$~\AA\, which we have studied previously\cite{Blunt2008Science}, and 1,4-diphenyl-1,3-butadiyne-3,3",5,5"-tetracarboxylic acid (DPBDTC) (Fig.~1i), with $d_1=9.5$~\AA\, synthesised specifically for this study; $d_2=9.6$~\AA\ for both molecules.  Several closely related molecules have been studied previously  \cite{Lackinger2009,Zhou2007,Zhao2010,Blunt2008ChemCommun}, but do not display an extended random tiling, although frustrated crystallisation has been observed\cite{Zhou2007}. To tile the plane perfectly a rhombus must have internal angles of $60^{\circ}$ and $120^{\circ}$, a lozenge \cite{Wilson2004}, for which $d_1=d_2$.

\begin{figure*}
\centering
\includegraphics[width=150mm]{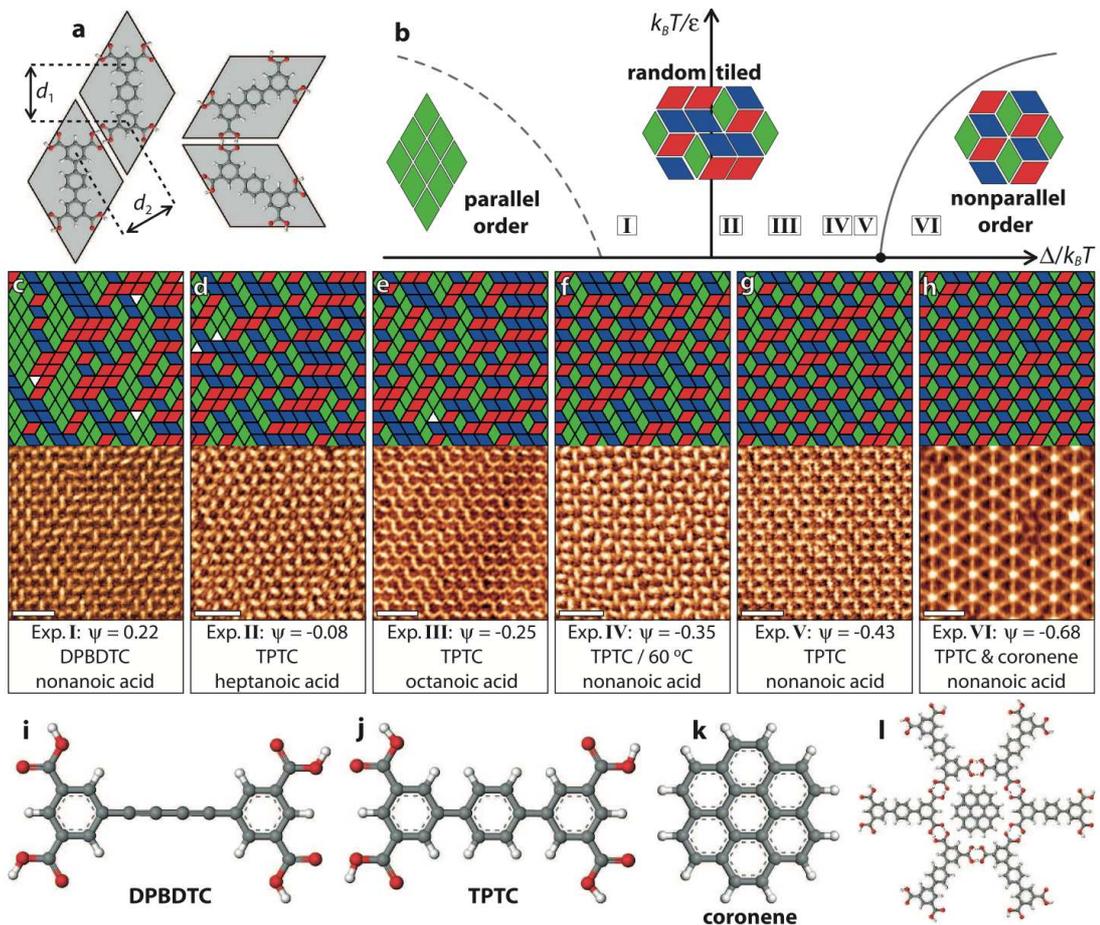}
\caption{{\bf Tetracarboxylic acid supramolecular assemblies and rhombus tilings.} {\bf a}, Parallel and non-parallel intermolecular bonding orientations, with backbone $d_1$, and bond, $d_2$, lengths indicated, overlaid onto the corresponding rhombus tile representations. {\bf b}, Schematic phase diagram of the interacting rhombus tiling model as predicted by theory. Three phases are expected depending on the interaction energy $\Delta$: a random tiled phase with critical correlations for small $|\Delta|$ (a so-called ``Coulomb'' phase), an ordered phase dominated by non-parallel bonding at large positive $\Delta$, and a second ordered phase dominated by parallel bonding at large negative $\Delta$.  The transition between the random tiled and non-parallel phases is expected to be continuous at $T>0$ (indicated by a full line in the sketch), and that to the parallel phase first-order (dashed line). {\bf c-h}, STM images (sections of larger scans) of tetracarboxylic acid supramolecular networks at alkanoic-acid-HOPG interfaces and the corresponding rhombus tilings, in order of decreasing $\psi$. Experiments are labelled I-IV and are performed at room temperature (apart from IV which was at 60$^o$C) using different combinations of TPTC, DPBDTC, coronene and solvents as specified in text boxes. STM image contrast originates from molecular backbones (and, in {\bf h}, coronene). All scale bars 50~\AA, see Methods for imaging parameters. {\bf i-k}, Molecular ball and stick diagrams of; {\bf i}, DPBDTC; {\bf j}, TPTC; and, {\bf k}, coronene. {\bf l}, Diagram of coronene adsorbed at the vertex of six TPTC molecules.}
\end{figure*}
A random tiling is considered {\em ideal} when the energies of the non-parallel and parallel alignments, given by $\varepsilon_N$ and $\varepsilon_P$ respectively, are degenerate, $\varepsilon_N = \varepsilon_P$  \cite{Alet2006, Fradkin2007, Castelnovo2007, Jacobsen2009}. In this case the tiling configuration is determined solely by maximising configurational entropy.  Such tilings possess no translational order and display correlations in tile orientation which have a logarithmic dependence on spatial separation, a signature of a `Coulomb' phase \cite{Henley1999}. However, if there is an energetic asymmetry, ${\Delta \equiv \varepsilon_N-\varepsilon_P \neq 0}$, the degeneracy between the exponentially large number of tilings is broken. In this case an additional internal energy contribution to the free energy competes with the entropic term (rhombus tiles with asymmetric interactions are described as {\em interacting} in many theoretical models\cite{Alet2006, Fradkin2007, Castelnovo2007, Jacobsen2009}). For small ${\Delta}$, random tiling phases (with properties which vary continuously with $\Delta$) are expected, but large ${\Delta}$ values are predicted to lead to transitions to ordered phases \cite{Alet2006, Fradkin2007, Castelnovo2007, Jacobsen2009}.  The parameter ${\Delta}$ may be considered to be analogous to the coupling constant $J$ which appears in Ising models of magnetism and leads to ordered phases with broken symmetry for $\vert J \vert \gtrsim k_BT$, which may be either ferromagnetic or antiferromagnetic depending on the sign of $J$, while entropic terms in the free energy dominate for $\vert J \vert << k_BT$ in which case a paramagnetic phase occurs. Our experiments and simulations show that this rich phase behaviour may be investigated by preparing molecular arrays with differing values of $\Delta$.

Figures~1c to 1h show images of molecular networks prepared under several different conditions (see Methods), referred to as Experiments I to VI, respectively, and the corresponding rhombus tiling representations, where each molecule is represented by a rhombus coloured according to its orientation. To characterise our experimental tilings we define an order parameter, ${\psi=(n_0p-p_0n)/(n_0p+p_0n)}$, where $n$ and $p$ represent the fraction of rhombus tile junctions in non-parallel and parallel orientations, respectively, and ${n_0 \simeq 0.608}$ and ${p_0 \simeq 0.392}$ are the equivalent values for a defect-free, ideal, random tiling and are estimated numerically (see Methods). As such, $\psi = 1$ in a fully parallel phase, ${\psi = -1}$ in a fully non-parallel phase, and ${\psi=0}$ for an ideal tiling. $\psi$ is calculated for each Experiment (I to VI), and the images (Figs.~1c  to 1h) are placed in order of decreasing $\psi$, spanning ${\psi=0.22}$ to ${\psi=-0.68}$. 

Fig.~1b is a schematic of the expected equilibrium phase diagram\cite{Alet2006, Fradkin2007, Castelnovo2007, Jacobsen2009}. There are two relevant thermodynamic parameters: temperature, $T$, and energetic bias, $\Delta$. Ideal random tilings are observed for $\Delta=0$. For $\Delta>0$ non-parallel bonding is favoured, resulting in increasingly negative values of $\psi$. For sufficiently large $\Delta$ the system orders into a crystalline phase dominated by non-parallel bonds. This transition occurs at $\Delta/k_BT = 0.454(3)$\cite{Jacobsen2009,footnote}. For $\Delta<0$ parallel bonding is favoured, resulting in increasingly positive values of $\psi$. The random tiled phase extends until $\Delta/k_BT \approx -0.3$ where the system undergoes a first-order transition\cite{Castelnovo2007} to a crystalline phase dominated by non-parallel bonds. We re-iterate the analogy with magnetic systems where transitions from disordered to ordered phases occur as the absolute value of the coupling constant is increased beyond some critical value.

It is clear from Fig.~1 that different experimental conditions give rise to tilings with varying degrees of order. In Experiment I we investigated DPBDTC in nonanoic acid (Fig.~1c). For this arrangement there are more parallel bonds than expected for an ideal random tiling and $\psi=0.22$. A solution of TPTC in heptanoic acid (Exp. II, Fig.~1d) results in a molecular tiling with $\psi=-0.08$. This is the molecular tiling closest to ideal (i.e. $\psi=0$) of all the systems we have studied. For TPTC a change of solvent \cite{Kampschulte2006,Tahara2006} from heptanoic acid to octanoic (Exp. III, Fig.~1e, $\psi=-0.25$) or to nonanoic (Exp. V, Fig.~1g, $\psi=-0.43$) acid, solvent molecules with similar chemistry but slightly larger size, leads to a progressive increase in the fraction of non-parallel bonds. Experiment IV is performed with TPTC and nonanoic acid at an elevated temperature of 60~$^{\circ}$C (Fig.~1f, $\psi=-0.35$). Comparing this molecular tiling with the equivalent tiling prepared at 19~$^{\circ}$C (room temperature) we see that the increase in temperature leads, as expected, to a value of $\psi$ closer to zero. Finally, the most negative order parameter, $\psi=-0.68$, is observed when coronene (Fig.~1k) is added to a solution of TPTC and nonanoic acid (Exp. VI, Fig.~1h). Coronene becomes incorporated in the molecular network in a site which enhances the stabilisation of the non-parallel arrangement (Fig.~1l), a capture process consistent with previous studies of coronene-carboxylic acid and other systems \cite{Blunt2008ChemCommun,Furukawa2007,Griessl2004,Wu2007}.

These results show that small modifications in experimental parameters lead to the exploration of the rhombus tiling model phase space. To determine the placement of the experimental structures on the phase diagram in Fig.~1b, we have undertaken extensive numerical simulations of rhombus tilings. The details of the numerical scheme are provided in the Methods section and represent a generalisation of previous work\cite{Garrahan2009, Stannard2010} to the case $\Delta\neq0$. Fig.~2a shows the numerically computed order parameter $\psi$ as a function of scaled bias $\Delta/k_BT$. Vertical dashed and solid lines indicate where transitions from random to ordered phases occur. The six experiments are marked on the numerical curve according to their measured $\psi$ values. Experiments II to V span the random tiled region with negative $\psi$; Experiment VI lies beyond the phase boundary in the non-parallel phase; Experiment I is close to the phase boundary with the ordered parallel phase. 

\begin{figure}
\centering
\includegraphics[width=77mm]{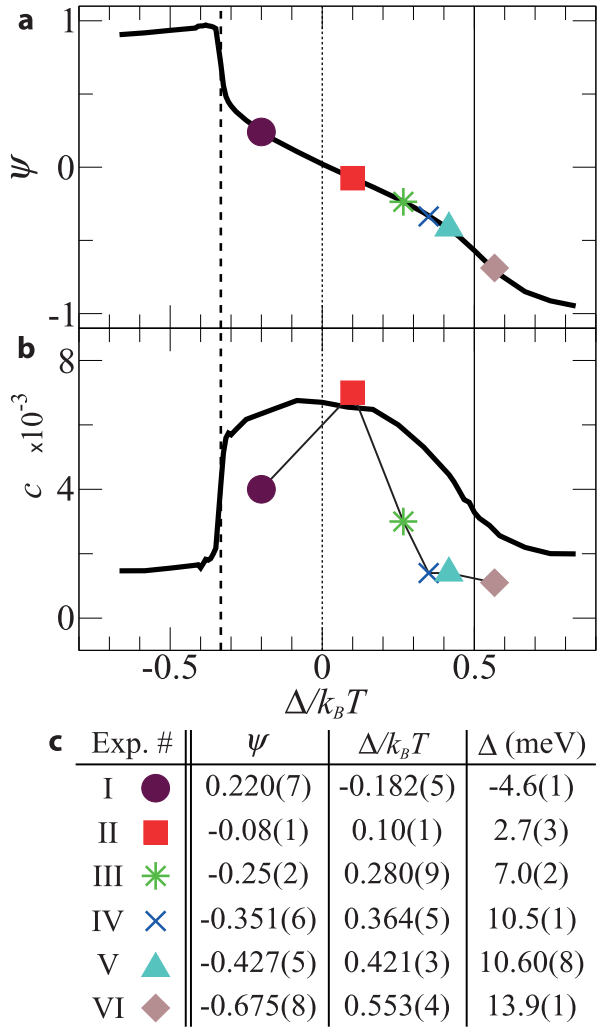} 
\caption{{\bf Comparison of experiments to simulations of the rhombus tiling model.} {\bf a}, Order parameter $\psi$ as a function of scaled bias $\Delta / k_B T$.  The full curve is the result of numerical simulations of the rhombus tiling model (see Methods) for $\varepsilon_N / k_B T = 5/3$ (at this value the concentration of defects $c$ is comparable to that observed experimentally).   The vertical lines indicate the approximate location of the continuous transition to the non-parallel ordered phase, $\Delta / k_B T = 0.49(2)$ (slightly larger than that calculated\cite{Jacobsen2009} for $T = 0$ - see Methods and footnote\cite{footnote}), and the first-order transition to the parallel phase, $\Delta / k_B T = -0.34(3)$.  The experimental points are placed on the curve according to the measured values of $\psi$. {\bf b}, Concentration of defects as a function of bias.  The full curve is the numerical result.  The abscissa values for experimentally measured defect concentrations are fixed by panel {\bf a}: the experiments show the predicted non-monotonic dependence on bias. {\bf c}, Estimate of bias $\Delta$ for the six experiments.  Notice that $\Delta$ for Experiments IV and V, which correspond to the same system (TPTC in nonanoic acid) at different temperatures, is the same within experimental error, as expected.}
\end{figure}

The numerical results in Fig.\ 2a provide a relationship between the order parameter $\psi$ and $\Delta/k_BT$ which allow an estimation of the latter for each of our experiments. Furthermore, using the known temperature at which the experiments were performed we can estimate values of $\Delta$, as tabulated in Fig.~2c. A comparison of Experiments IV and V is of particular interest: both were performed using TPTC in nonanoic acid but with molecular networks prepared at different temperatures. Accordingly we would anticipate the same value of $\Delta$ and we find excellent agreement between the values inferred from simulations. This result provides strong support for our analysis and confirmation that the molecular arrays can be understood in terms of equilibrium structures. From our experiments it is clear that measurable variations in the tiling statistics may be identified even for variations of $\Delta$ of $\lesssim 5$~meV, much less than the thermal energy at room temperature. 

A second experimental measure, which can also be compared with numerical results, is the concentration of defects. These `half-rhombus' vacancies are topological defects and are present in several of the images shown in Fig.~1. For an ideal tiling their concentration, $c$, has been shown to be temperature dependent \cite{Garrahan2009} (we choose an effective temperature in our simulations to match, approximately, the maximum defect density observed in experiments) and we show in Fig.~2b that $c$ is also controlled by $\Delta$. We highlight the non-monotonic dependence on bias which results from the strong suppression of isolated triangular defects in the ordered phases. The experimentally-determined defect data is plotted against the theoretical curve using the value of $\Delta/k_B T$ obtained from Fig.~2a. We note that the non-monotonic behaviour and suppression of defect formation in the ordered phases is reproduced in our data thus providing further support of the validity of our approach to determine the value of  $\Delta$ for our experiments.

\begin{figure*}
\centering
\includegraphics[width=170mm]{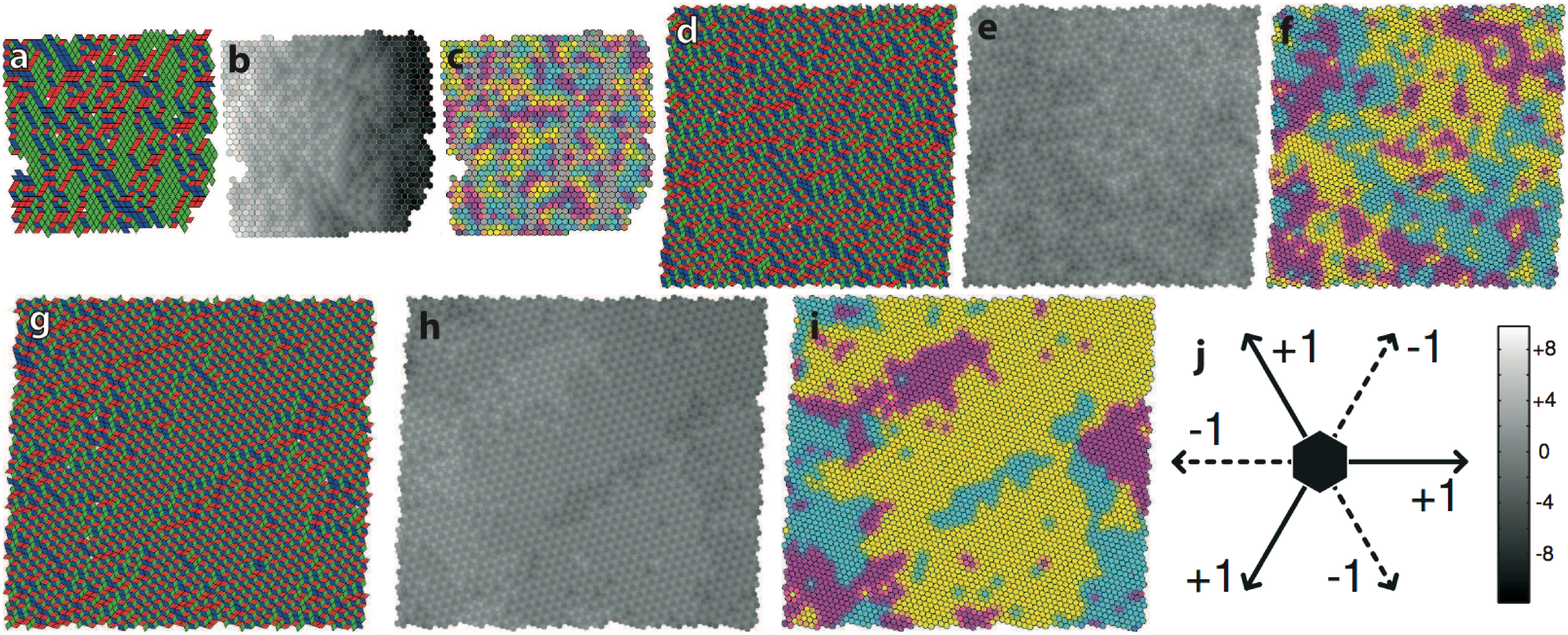}
\caption{{\bf Ordered phase analysis of tetracarboxylic acid supramolecular networks.} {\bf a-c}, Experiment I (DPBDTC in nonanoic acid): ({\bf a}) tiling representation of a 50~nm STM image; ({\bf b}) height field representation, showing that at this scale the system is `tilted'; ({\bf c}) non-parallel phase domain representation (we colour cyan/yellow/magenta the three possible different sub-lattice arrangements of the non-parallel ordered phase). {\bf d-f}, Experiment IV (TPTC in nonanoic acid): ({\bf d}) tiling representation (80~nm image),  ({\bf e}) height field representation; ({\bf f}) non-parallel phase domain representation. {\bf e} and {\bf f} show that the structures of Experiment IV are neither tilted, nor display any significant non-parallel phase order, as expected of a random tiling.   {\bf g-i}, Experiment VI (TPTC and coronene): ({\bf g}) tiling representation (100~nm image); ({\bf h}) height field representation; ({\bf i}) non-parallel ordered domain map. While {\bf h} shows that the structure is overall flat, {\bf i} shows that one ordered non-parallel domain has become prevalent indicating a spontaneous breaking of symmetry due to a transition to the ordered non-parallel phase. {\bf j}, Scheme for calculating the height field: moving along the edges of tiles the field $h(x,y)$ increases/decreases by the amounts indicated. The bar shows the scaling of height to contrast in the height representations; the same scaling is used in (b), (e) and (h).}
\end{figure*}

Rhombus tilings are commonly analysed as a projection of the interface of a simple cubic lattice onto a (111) plane \cite{Henley1999}. In this approach each vertex is assigned an effective height $h(x,y)$ ($x$ and $y$ are the co-ordinates of the vertex) using a simple procedure (outlined in Fig.~3j). An analysis of our images using this approach shows explicitly the difference between random and ordered phases as shown in Fig.~3.  Here we show larger area images for Experiments I, IV and VI, in three different representations.  The first is the tiling representation, as in Fig.~1 (Figs.~3a, 3d, and 3g). The second representation is given by the height map $h(x,y)$ (Figs.~3b, 3e, and 3h): Experiment I shows a clear height gradient, Fig.~3b, indicative of its closeness to the phase boundary to the parallel phase; in contrast the height maps for Experiments IV and VI, Figs.~3e and 3h, show averaged height gradients which are close to zero. This is expected for both the random and non-parallel ordered phase although these can be distinguished by the greater height fluctuations present for Experiment IV consistent with a random tiling phase.  

The third representation illustrates the spatial variation of non-parallel order.  In a perfectly non-parallel configuration the unit cell (e.g. see schematic inset of Fig.\ 1b) can be centred at any of the three sub-lattices of the triangular lattice.  Choosing a sublattice is the symmetry breaking associated with the transition to the non-parallel phase and in general we would expect to observe domains of non-parallel order corresponding to each of these three possible sub-lattices. Again a magnetic analogue is useful; in a ferromagnetic system domains of spin-up and spin-down magnetisation are formed whose size diverges as the system undergoes a transition from a paramagnetic to a ferromagnetic phase. For the case considered here there are three equivalent non-parallel ordered states into which the system can crystallise, rather than the two states available for a simple magnetic system.   In Figs.~3c, 3f, and 3i regions of non-parallel aligned tiles are coloured cyan/yellow/magenta depending on their assignment to one of the three possible non-parallel phases. For Experiment I, Fig.~3c, these domains are relatively small since the parallel tile alignment predominates. The domains are larger but finite for Experiment IV, Figs.~3f, in the random tiling phase.  For Experiment VI, Fig.~3i, one domain has become dominant and is comparable with the size of the system, indicative of a transition to the non-parallel phase.

Our results show conclusively that the phase space of two-dimensional rhombus tiles, which arises from asymmetric interactions, may be explored experimentally through the structural characterisation of molecular arrays. The changes in energy which give rise to different ordering are small compared to the thermal energy, an essential requirement for systems where entropy is significant. The dependence of the characteristic energetic bias, $\Delta$, on solvent indicates that the effective energetic difference between parallel and non-parallel molecular pairs adsorbed on graphite cannot be simply described by the interactions between two tetracarboxylic acid molecules and must include other more complex contributions. We highlight two possible influences of the solvent molecules on the effective interaction between adsorbed molecules.  Firstly the co-adsorption of solvent molecules on the graphite surface is possible\cite{Gutzler2010}, and expected, since the the network of tetracarboxylic acid molecules results in an array of nanopores which can accommodate guest molecules. As shown in a recent paper\cite{Blunt2011NatChem} these nanopores are hexagonal and may be classified into five inequivalent types which are formed from different combinations of parallel and non-parallel bonded molecular pairs. There is a significant variation of the capture of guest molecules in pores of different type\cite{Blunt2011NatChem}, and we therefore also expect variations in the stabilisation energies of the co-adsorption of solvent molecules in different pore types. Since the energy of co-adsorbed molecules makes a contribution to the overall stability\cite{Gutzler2010} of a given pore, this effect provides a mechanism to break the degeneracy of the different pore types. Since different pores are formed by different combinations of parallel and non-paralell pairs, variations in guest inclusion stability will result in small effective energetic differences between the parallel and non-parallel molecular orientations.  A second consideration is that the solvents provide an effective medium with varying dielectric constant, as pointed out for these particular molecules by Kampschulte {\it et.al.}\cite{Kampschulte2006}, which would give rise to a solvent dependence of any energetic imbalance mediated, for example, by van der Waals interactions, or hydrogen-bonding, between neighbouring molecules (see Cook et.al.\cite{Cook2007} for a discussion of solvent effects on intermolecular interactions in a solution environment).

A complete microscopic description of energy differences on this scale are beyond the confidence levels normally associated with modern numerical approaches to molecular modelling (e.g. density functional theory, molecular dynamics). Furthermore the inclusion of solvent molecules makes a detailed microscopic description and prediction of energies particularly problematic. However, our experiments give clear limits on the tolerance of the random tiling phase to symmetry breaking - a difference in binding energies of $\gtrsim10$meV will drive the structures into an ordered phase. 

Our study demonstrates that prototypical self-assembled systems such as supramolecular networks can reveal a richness of exotic phases which have been studied widely by theorists and are analogous to those found in complex condensed-matter systems. In addition our work highlights the role of entropy in the balance between order and disorder in molecular templates.

\section*{Methods}
{\bf Chemicals.} 1,4-diphenyl-1,3-butadiyne-3,3",5,5"-tetracarboxylic acid (DPBDTC) and  {\it p}-terphenyl-3,5,3",5"-tetracarboxylic acid (TPTC) were synthesized in-house, details of their synthesis can be found in the accompanying Supplementary Material and Ref.\ \onlinecite{Blunt2008Science} respectively. Other chemicals used were coronene ($\geq$95.0\%, Fluka), heptanoic acid ($\geq$97\%, Sigma), octanoic acid ($\geq$98\%, Aldrich), and nonanoic acid ($\geq$95\%, Fluka).

{\bf STM experiments.} STM images were acquired with an Agilent Technologies 4500 PicoPlus STM using a PicoScan controller and STM tips formed from mechanically cut PtIr (80:20) wire. Saturated solutions of the desired adsorbate (DPBDTC or TPTC) were prepared by placing an excess of solid in the desired solvent (heptanoic, octanoic, or nonanoic acid). For Experiment VI, the solution used combined equal volumes of a saturated solution of TPTC in nonanoic acid and coronene in nonanoic acid (conc. $1.5\times10^{-4}$~mg/mL). All solutions were ultrasonically agitated to ensure complete dissolution. To form the supramolecular networks, a 10~$\mu$L droplet of solution was deposited onto a mechanically cleaved HOPG substrate. Imaging at the alkanoic-acid-HOPG interface commenced immediately after approach of the STM tip. For Experiment IV, the substrate was heated to 60~$^{\circ}$C prior to solution deposition and held at this temperature for the duration of the experiment. Figure~1 STM imaging parameters ($V_{tip}$/$I$): 1c, +1.1~V/10~pA; 1d, +1.2~V/15~pA; 1e, +1.2~V/5~pA; 1f, +1.25~V/12~pA; 1g, +1.0~V/13.5~pA; 1h, +1.0~V/30~pA.

{\bf Numerical simulations.} The rhombus tilings were simulated using the same model as in Ref.\ \onlinecite{Garrahan2009} supplemented by interactions which varied depending on whether neighbouring tiles were parallel (binding energy $\epsilon_P$) or non-parallel (binding energy $\varepsilon_N$).  Simulations were performed with continuous time Monte Carlo which allows to access very long times at low temperatures  (i.e. very low number of defects) on systems sizes between $N=18^2$ to $99^2$ and averaged over $10^3$ samples for each state point.  The location of the continuous transition from the random tiled to the ordered non-parallel phase at $T \neq 0$ (the location at $T=0$ is known exactly\cite{Jacobsen2009}) was determined from finite size scaling analysis of susceptibilities and higher order cumulants of $\psi$ and of the `columnar' order parameter (difference in density of non-parallel order in each sublattice) as in Ref. \onlinecite{Alet2006}; the location of the first-order transition to the ordered parallel phase from the jump in $\psi$ and the `tilt' order parameter (difference in density of each kind of tile), cf. Ref.\ \onlinecite{Castelnovo2007}.

\section*{Acknowledgements}
We are grateful to the UK Engineering and Physical Sciences Research Council (EPSRC) for financial support under grant EP/D048761/01. JPG was supported by EPSRC Grant No. GR/S54074/01. AS was supported by the Leverhulme Trust [ECF/2010/0380] and the EPSRC [EP/P502632/1]. MS acknowledges receipt of a Royal Society Wolfson Merit Award and an ERC Advanced Grant. NRC gratefully acknowledges receipt of a Royal Society Leverhulme Trust Senior Research Fellowship.

\end{document}